\documentclass[
  prb,
  twocolumn,
  floatfix,
  superscriptaddress,
  showpacs]
  {revtex4}

\usepackage{graphicx}

\newcommand{\e}{\text{e}}
\newcommand{\kB}{k_{\text{B}}}
\newcommand{\rhorh}{\rho_{\text{rh}}}

\begin{document}

\title{Cluster model of decagonal tilings}

\author{Michael Reichert}
\altaffiliation[Present address: ]{Fachbereich Physik, Universit\"at Konstanz,
D-78457 Konstanz, Germany}
\email{michael.reichert@uni-konstanz.de}
\affiliation{Institut f\"ur Theoretische und Angewandte Physik,
Universit\"at Stuttgart, \\ D-70550 Stuttgart, Germany}
\affiliation{Fachbereich Physik, Universit\"at Konstanz,
D-78457 Konstanz, Germany}
\author{Franz G\"ahler}
\affiliation{Institut f\"ur Theoretische und Angewandte Physik,
Universit\"at Stuttgart, \\ D-70550 Stuttgart, Germany}

\date{\today}

\begin{abstract}
A relaxed version of Gummelt's covering rules for the aperiodic decagon
is considered, which produces certain random-tiling-type structures.
These structures are precisely characterized, along with their 
relationships to various other random tiling ensembles. The relaxed 
covering rule has a natural realization in terms of a vertex cluster 
in the Penrose pentagon tiling. Using Monte Carlo simulations, it is
shown that the structures obtained by maximizing the density of this 
cluster are the same as those produced by the corresponding covering 
rules. The entropy density of the covering ensemble is determined 
using the entropic sampling algorithm. If the model is extended by 
an additional coupling between neighboring clusters, perfectly ordered 
structures are obtained, like those produced by Gummelt's perfect 
covering rules.
\end{abstract}

\pacs{61.44.Br, 64.60.Cn}

\maketitle

\section{Introduction}
\label{sec_intro}

\begin{figure}[b]
\centerline{\includegraphics[width=\columnwidth]{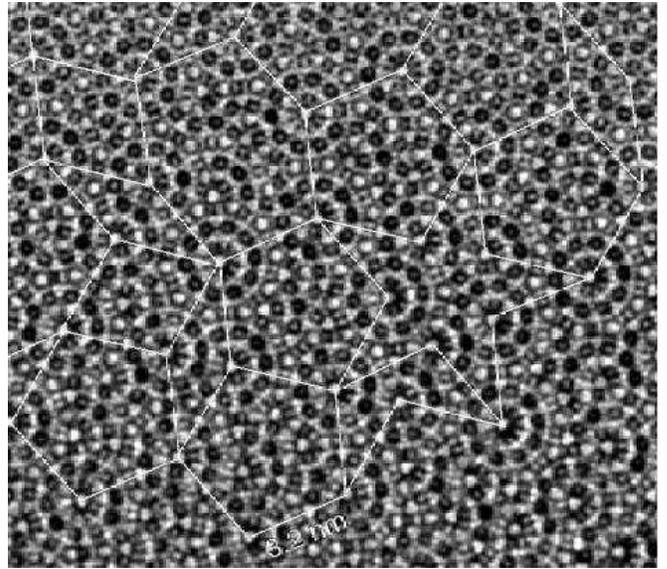}}
\caption{High-resolution transmission electron microscopy image of 
decagonal Al-Ni-Co (courtesy of S.~Ritsch and C.~Beeli, compare 
Ref.~\onlinecite{rit96}). The superimposed tiling has been reconstructed
by an automated procedure.\cite{sol01}}
\label{fig_em-picture}
\end{figure}

Many quasicrystals are completely covered by overlapping copies of a
single cluster (Fig.~\ref{fig_em-picture}). Two overlapping clusters
must agree in the overlap 
region, which restricts the possible relative positions and
orientations of neighboring clusters. Cluster overlaps therefore
create order, in favorable  cases even perfect quasiperiodic
order.

This observation can be used to formulate several variants of
an ordering principle for quasicrystals (for a review, see 
Ref.~\onlinecite{gae00}): A perfect quasicrystal can be obtained by requiring
either that a given cluster completely covers the structure, or that
the cluster has maximal density in the structure, or that it covers 
the structure with maximal density. With such ordering principles, 
perfect quasiperiodic order could be obtained for
decagonal,\cite{gum96,jeo97} octagonal,\cite{ben99} and dodecagonal 
\cite{gae01} tilings and quasicrystals. Assuming that such a cluster is an
energetically preferred atomic configuration, the maximization of the
cluster density minimizes the free energy. With this hypothesis, the
covering approach might serve as a simple thermodynamic mechanism 
for the formation of quasicrystals. The covering approach can be 
regarded as a particularly simple realization of energy based 
matching rules, where only the most important local configurations
(the clusters) have to be preferred energetically \cite{jeo94}, 
not all allowed local configurations. As a variant, it has also been
suggested\cite{hen97} to penalize the worst local configurations,
instead of prefering the best ones, which provides another way to
simplify the matching rule approach.

The same ordering principles can also be used to produce supertile
random tiling structures, which are locally ordered but show disorder
on larger scales. These structures are obtained 
whenever the chosen cluster is not selective enough and hence
allows too many different overlaps.\cite{gae00} This
happens, in particular, if the cluster is too small to restrict the
number of different overlaps,\cite{gae95} or if it is too
symmetric. In this respect, it is interesting to note that {\it
asymmetric} clusters seem to be preferred by the electronic structure
in decagonal quasicrystals.\cite{yan01} Whereas supertile random tilings 
are locally ordered for energetic reasons, their long range order
is produced by entropy maximization, as is the case for other random 
tilings\cite{hen91}.

In this paper, we will concentrate on cluster models for decagonal
structures. Gummelt's aperiodic decagon 
\cite{gum96} provides a striking example how perfect quasiperiodic
order can be obtained by a simple cluster covering principle. This
example has been so convincing, that many researchers tried
to map their experimental structures to the Gummelt decagon,
even though the fit in the overlap was often not perfect (see, e.~g., 
Refs.~\onlinecite{yan01} and \onlinecite{abe00}), and the overlapping
constraints not exactly equivalent. 
However, many experimental decagonal quasicrystal structures are
not perfectly quasiperiodic, and it is therefore interesting to consider
also overlap rules which are less restrictive than the perfect rules
of Gummelt, and which do not enforce perfectly ordered,
but rather (supertile) random tiling structures. 

The analysis of such relaxed overlap rules and their corresponding
structures will be the main topic of this paper.\cite{sendai}
In Sec.~\ref{sec_covering},
two different relaxed versions of Gummelt's overlap rules are discussed,
and the structures which they produce are precisely characterized, along with
their relationships to various other random tiling
ensembles. Subsequently, in Sec.~\ref{sec_clustermax} we introduce
a vertex cluster in the Penrose pentagon tiling (PPT) whose structure
imposes the previously discussed overlap constraints in a natural
way. It is shown by Monte Carlo (MC) simulations that the structures with
maximal density of this cluster are the same as those produced
by the corresponding overlap rules. In Sec.~\ref{sec_entropy}, we
determine the entropy density of the set of states with maximal
cluster density, using the entropic sampling algorithm. 
An additional coupling between neighboring clusters is introduced
in Sec.~\ref{sec_coupling}, and it is shown that this coupling
is capable of ordering the random tilings to perfectly ordered structures.

\section{Coverings for perfect and \protect\\ random Penrose pentagon tilings}
\label{sec_covering}

\begin{figure}[b]
\centerline{\includegraphics[width=\columnwidth]{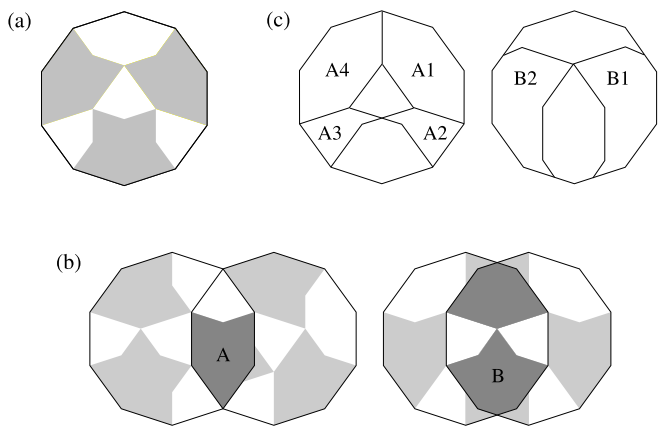}}
\caption{Gummelt decagon (a), representative A- and B-overlap (b),
and allowed overlap zones (c).}
\label{fig_decagon}
\end{figure}

It is well known that each covering of the plane by Gummelt's aperiodic
decagon (Fig.~\ref{fig_decagon}(a)) is equivalent to a
perfect Penrose tiling,\cite{gum96} if the covering has the following
property: Whenever two decagons overlap, their colorings agree in the
entire overlap region. It turns out that Gummelt's rule permits only
two different types of overlaps, which are shown in
Fig.~\ref{fig_decagon}(b): the smaller A- and the larger
B-overlaps. Furthermore, due to the coloring, there are only certain 
overlap zones for allowed overlaps with neighboring decagons: four for
A- and two for B-overlaps (Fig.~\ref{fig_decagon}(c)). 
This altogether is what we will call the
{\it perfect rule} (in order to distinguish it from other variants
being discussed later).

\begin{figure}[t]
\centerline{\includegraphics[clip,width=\columnwidth]{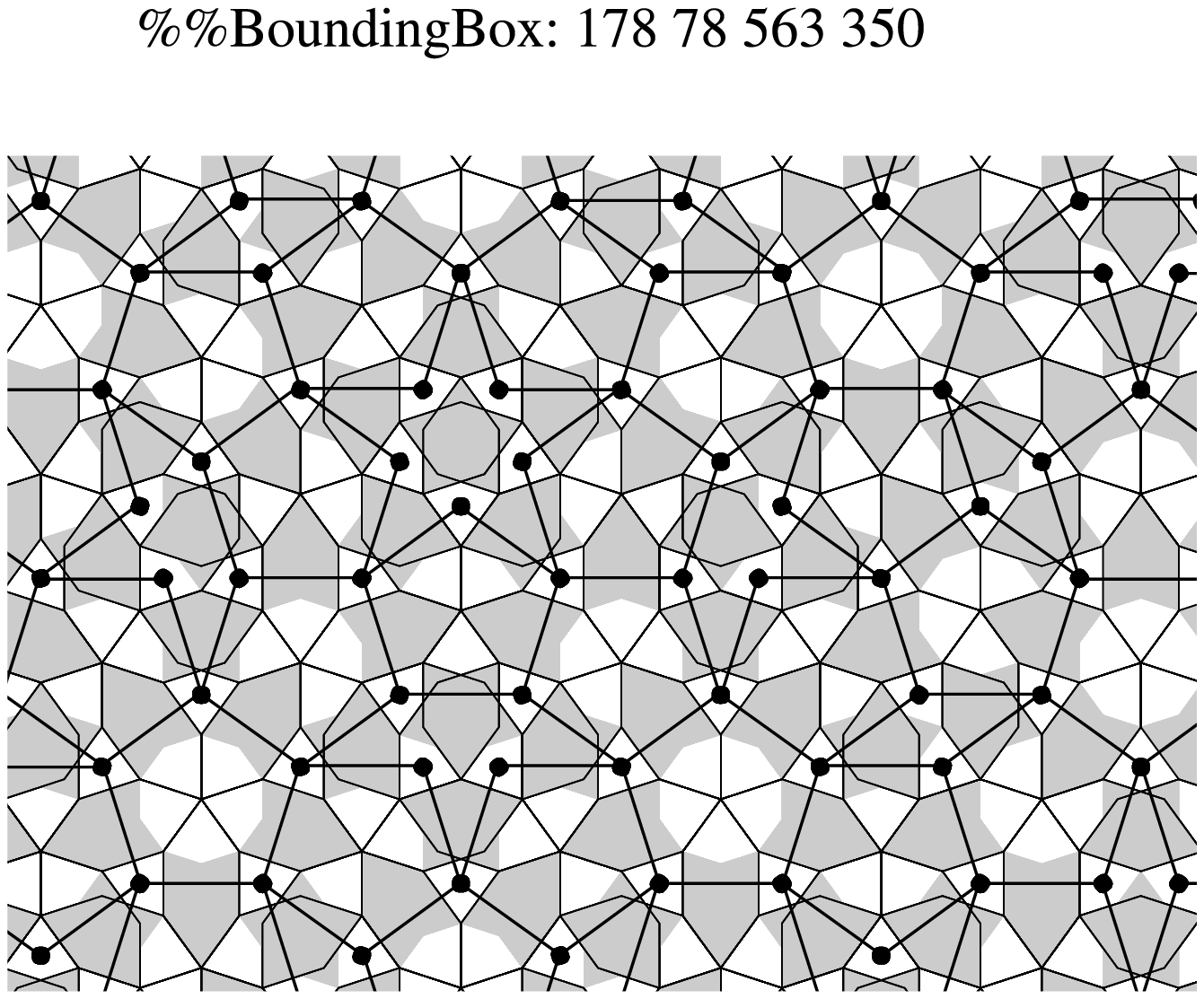}}
\caption{Perfect PPT, superimposed on the corresponding
Gummelt decagon covering.}
\label{fig_covering}
\end{figure}

The decagon centers of such a {\it perfect covering}
form the vertex set of a {\it perfect PPT} (Fig.~\ref{fig_covering}).
Conversely, each PPT can be obtained from exactly one covering
satisfying the Gummelt overlap rules. We therefore have a local
one-to-one correspondence between PPTs and Gummelt coverings. 
As the Gummelt decagon represents a cluster in the corresponding
quasicrystal, we will often use the term cluster for the covering decagon.

In order to allow for partially disordered coverings, Gummelt et 
al.\cite{gum00,gum02} have proposed to relax the overlap rules to
some extent. 
To understand the type of relaxation, recall that if the perfect rules 
are obeyed, a decagon may have small A-overlaps with neighboring decagons 
in four possible directions, and bigger B-overlaps with neighboring decagons 
in two possible directions (Fig.~\ref{fig_decagon}(c)). 
The coloring in the overlap region 
has an orientation, which must be respected. All possible overlaps are 
therefore {\it oriented}. As a relaxation of the perfect rule, Gummelt 
et al.\cite{gum00,gum02} have proposed to abandon this orientation 
constraint, and to retain only the {\it non-oriented overlap zones}, 
as shown in Fig.~\ref{fig_decagon}(c). This overlap rule will be
referred to as the fully relaxed rule.

There is a natural intermediate rule between the perfect and the
fully relaxed rule. In this variant, which will be called the 
{\it relaxed rule}, the orientation condition is abandoned only for the
small A-overlaps, but is retained for the larger B-overlaps. 
This kind of overlap rule can be motivated physically as follows:
The large B-overlaps result in a strong interaction between the two
overlapping clusters, which must be in its ground state, whereas the 
small A-overlaps only lead to a weak interaction with a small
energy difference between differently oriented A-overlaps. 
This intermediate rule and the resulting structures will be the 
main topic of this paper.

\begin{figure}[t]
\includegraphics[clip,width=\columnwidth]{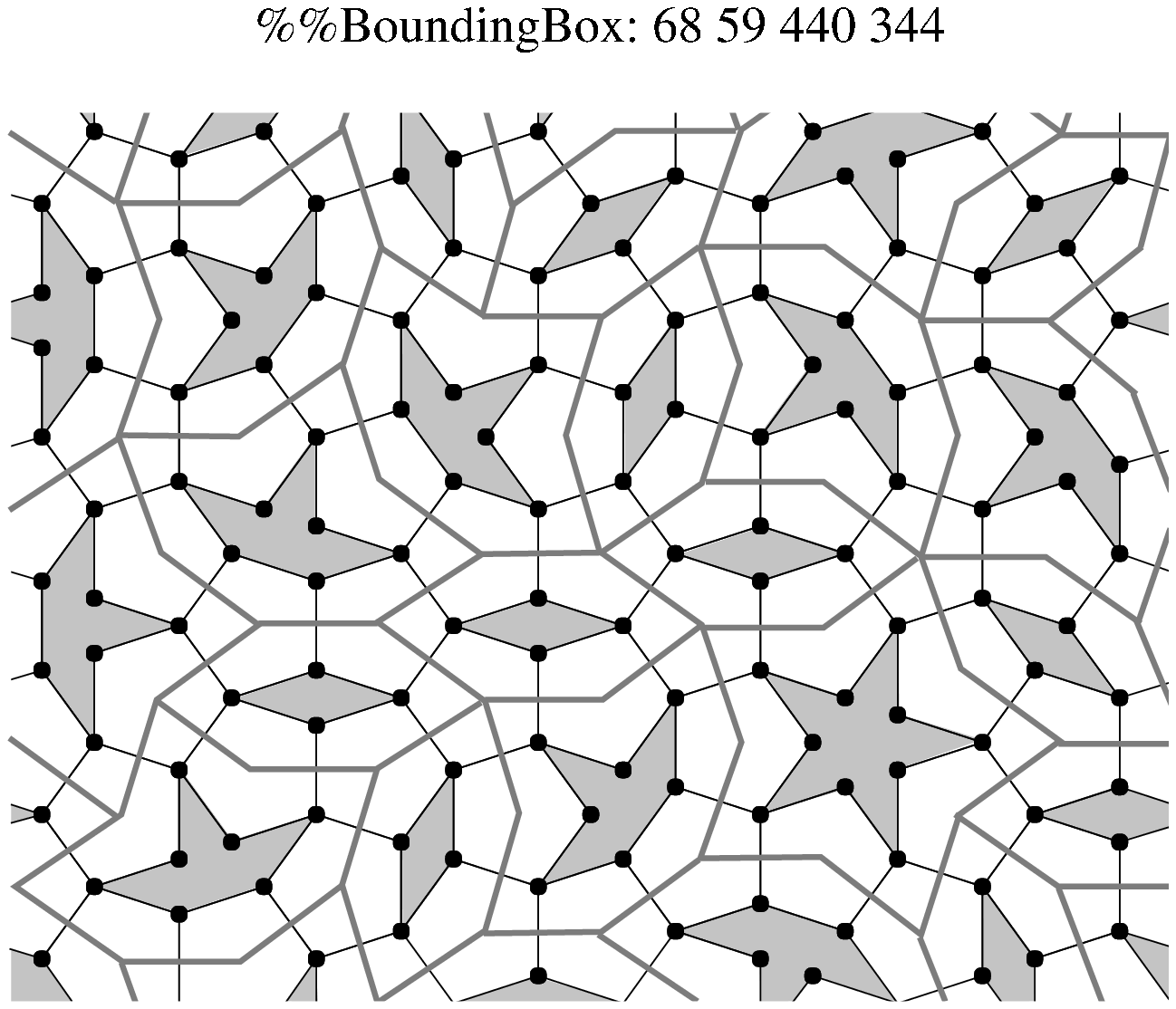}
\caption{Random PPT with all the spiky tiles (shaded in gray)
surrounded by pentagons. This tiling is equivalent to a random HBS
tiling (gray lines), whose tile edges connect the centers of
neighboring pentagons.}
\label{fig_hbs-tiling}
\end{figure}

Gummelt et al.\cite{gum00,gum02} have shown that each covering 
satisfying the fully relaxed rule has the property that its cluster 
centers form the vertex set of a random PPT. It has
the additional property that all the spiky tiles
(stars, ships, and rhombi; shaded in gray in Fig.~\ref{fig_hbs-tiling})
are completely surrounded by pentagons. 
(In the following, when we use the term ``random PPT'', we always mean one
satisfying  this extra condition; more general ones do not play any
role here.)  Such a random PPT is equivalent to a random 
hexagon-boat-star (HBS) tiling (gray lines in Fig.~\ref{fig_hbs-tiling}). 
Since coverings satisfying the more 
restrictive relaxed rule also satisfy the fully relaxed rule,
their cluster centers form the vertex set of a random
PPT, too. Conversely, it is easy to see that every
random PPT can arise both from relaxed and from
fully relaxed coverings. The only difference between relaxed and
fully relaxed coverings is the number of coverings associated with
a given random PPT.

\begin{figure}[b]
\includegraphics[clip,width=5.2cm]{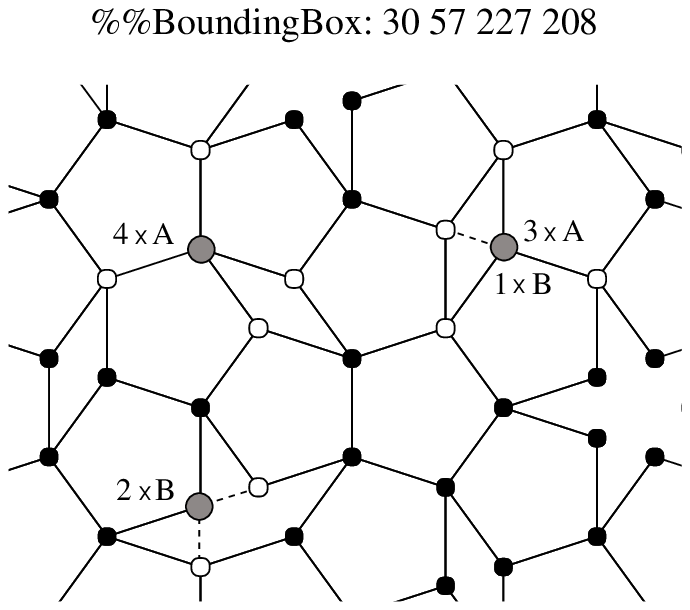}
\caption{The orientation of a decagon on a vertex in a PPT is fixed in
the case of four A- or two B-neighbors (left). The only vertices with
a choice for the decagon orientation are the obtuse rhombus corners, 
which have three A-neighbors and one B-neighbor (right).} 
\label{fig_orientations_1}
\end{figure}

\begin{figure}[b]
\includegraphics[width=\columnwidth]{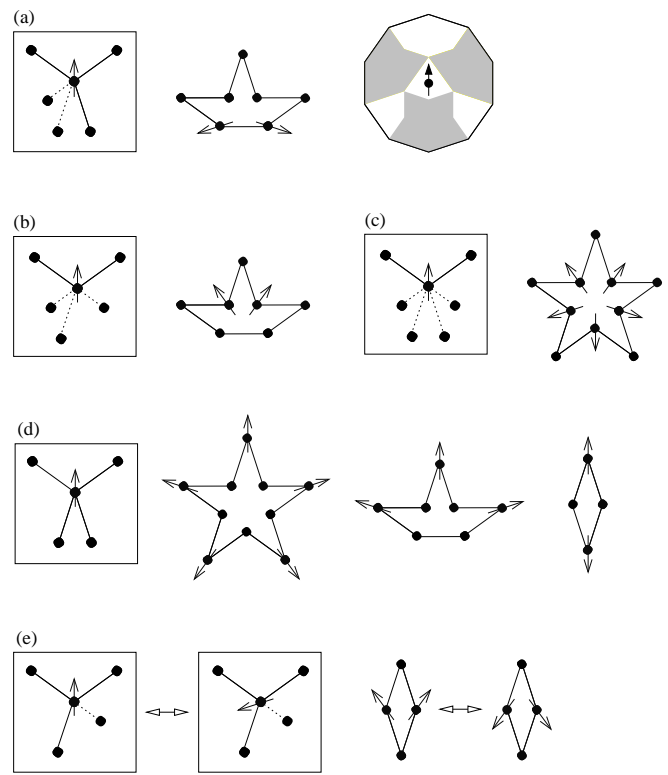}
\caption{Orientation of a decagon on a vertex in the PPT, depending
on the local environment. The orientations are given
by an arrow, as defined in the upper right corner.}
\label{fig_orientations_2}
\end{figure}

To see this, we note that the orientation of a cluster on a vertex in
a PPT is completely fixed by the presence of four A-neighbors or two
B-neighbors (Figs.~\ref{fig_orientations_1} and
\ref{fig_orientations_2}(a--d)),  as in these cases the four A- or the
two B-overlap zones, respectively, are completely
saturated. A-neighbors are separated by an edge of a  
tile or a long diagonal across a ship or star, whereas B-neighbors 
are separated by a short diagonal of a rhombus, ship or star. 
The only vertices whose cluster orientation
is not fixed by their local environment in the tiling are the obtuse corners 
of the rhombi, which have three A-neighbors and one B-neighbor
(Figs.~\ref{fig_orientations_1} and \ref{fig_orientations_2}(e)). 
Therefore, two cluster orientations are possible for each obtuse rhombus 
corner. For the fully relaxed rule, where no orientation conditions for the 
overlaps have to be obeyed, we thus have altogether four choices 
per rhombus for the cluster orientations.

However, for the relaxed overlap rule we have to obey the orientation
condition for the B-overlaps. It is easily shown that this condition
is always satisfied for B-overlaps across ships and stars.
In order to fulfill the orientation condition for B-overlaps across
rhombi, too, the orientations of the clusters on opposite obtuse
rhombus corners cannot be chosen independently. If for one of the corners
an orientation is chosen, the orientation of the other is already
fixed, i.~e., the condition is satisfied only for two of the four
possible combinations mentioned above. 

\begin{figure}[t]
\includegraphics[width=\columnwidth]{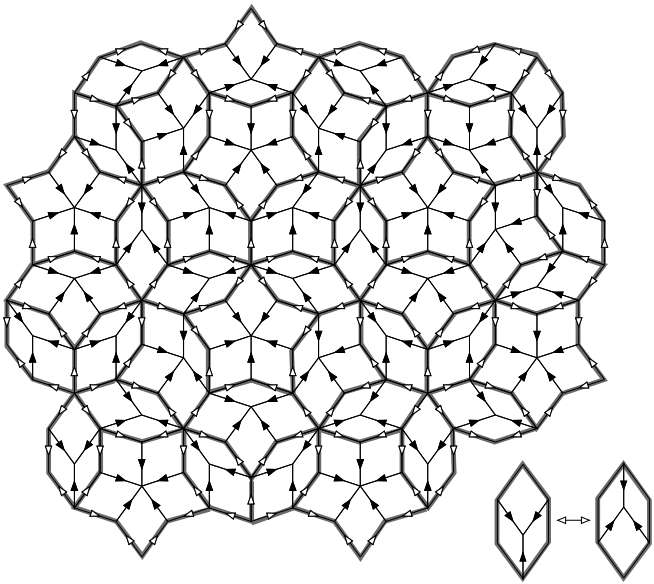}
\caption{Relationship between 4-level (thin black lines) and 2-level
random tilings (thick gray lines). The matching rules for the ``double'' 
arrows (here drawn in black) are still obeyed. Since the matching rules 
for the ``single'' arrows (here drawn in white) are no longer maintained, 
each hexagon of the 2-level or
random HBS tiling can be subdivided in two different ways, as shown in
the bottom right corner.}
\label{fig_2-level-tiling}
\end{figure}

In the same way, one can quantify the relationship between the cluster
coverings and certain variants of random Penrose rhombus tilings. The
random HBS tilings arise from random Penrose rhombus tilings still
satisfying the matching rules for ``double'' arrows (drawn in black in 
Fig.~\ref{fig_2-level-tiling}). Such random Penrose rhombus tilings
are also called 4-level random tilings.\cite{hen91,tan90} 
When the edges with a double arrow are
simply wiped out, we obtain the random HBS tilings, which are also
known as 2-level tilings.\cite{hen91} The relationship between
4-level and 2-level random tilings is not one-to-one: Whereas the
subdivision of boats and stars is unique, there are two choices for
the subdivision of each hexagon into rhombi
(Fig.~\ref{fig_2-level-tiling}), just as there are two possible
cluster assignments on the obtuse rhombus corners in the PPT, as
discussed above (Fig.~\ref{fig_orientations_2}(e)).
Since rhombi in the PPT and hexagons in the HBS tiling are in
one-to-one correspondence (Fig.~\ref{fig_hbs-tiling}), this implies
that the multiplicity of relaxed cluster coverings and 4-level random
tilings, related to a given random PPT, is the same. 
Apart from the extra multiplicities, the relaxed covering rule 
is therefore equivalent to the Penrose double-arrow matching rules
(ignoring single arrows), whereas the Gummelt convering rule is
equivalent to the full Penrose matching rules.

\section{Cluster density maximization}
\label{sec_clustermax}

In the last section, we have considered {\it cluster coverings},
where our clusters have simply been decagons with certain overlap rules.
Another variant of an ordering principle for quasicrystals based on a
cluster picture is the {\it cluster density maximization},%
\cite{gae00,jeo94,jeo97} which we will consider in the following.

\begin{figure}[b]
\includegraphics[width=\columnwidth]{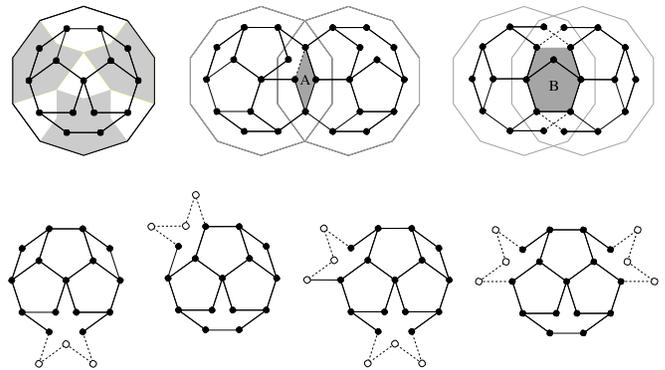}
\caption{Vertex cluster, superimposed on the Gummelt decagon (top left), and
representative A- and B-overlaps (top middle and right). This cluster
enforces the relaxed overlap rules. The bottom row shows examples
of tile configurations for the vertex cluster.}
\label{fig_cluster}
\end{figure}

The relaxed overlap rule discussed above allows for a very natural
realization in terms of a {\it vertex cluster} in the PPT. This vertex
cluster is shown in Fig.~\ref{fig_cluster}. We have to point out that
the tile edges are drawn only as a guide to the eye; they are not part
of the cluster, only the vertex set counts. It is easy to see that the
vertex set of the cluster cannot enforce the orientation of the small
A-overlaps of the Gummelt decagon, whereas the orientation of the
B-overlaps is intrinsically enforced. 
The A-overlap consists of a rhombus or a rhombic area inside a ship or
star without orientation, and the B-overlap is formed essentially of a
hexagon-shaped area with an interior vertex in asymmetric position
which yields an orientation (Fig.~\ref{fig_cluster}).

With this cluster, we can build a statistical model for the
cluster density maximization. We consider the set of all random PPTs
(we still require that spiky tiles are completely surrounded by
pentagons) and assign to each tiling a statistical weight which is
simply the number of vertex clusters it contains. With a suitable
MC algorithm, it is then possible to find the subensemble of
those random PPTs which have maximal cluster density. For this
purpose, we need a MC dynamics which is ergodic in the ensemble of all
random PPTs. By repeated flips it is then possible to turn any
random PPT into any other.

\begin{figure}[t]
\includegraphics[width=\columnwidth]{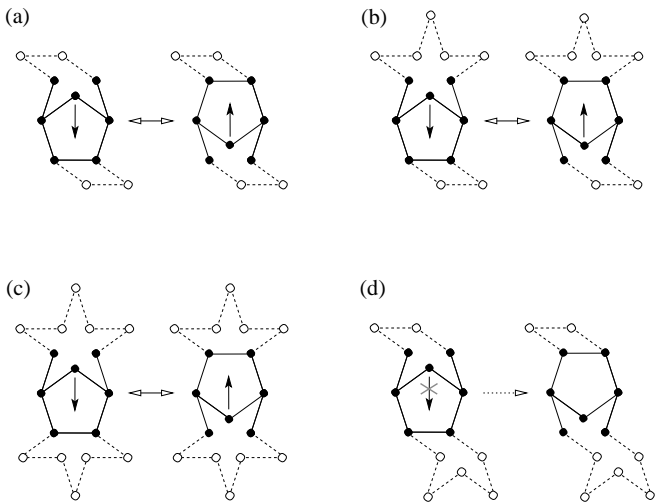}
\caption{Flip moves for the MC simulation. The flips (a--c) are
allowed, whereas move (d) is forbidden since it would produce a new
kind of tile (the zigzag-shaped configuration at the bottom).}
\label{fig_flips}
\end{figure}

\begin{figure}[b]
\includegraphics[width=\columnwidth]{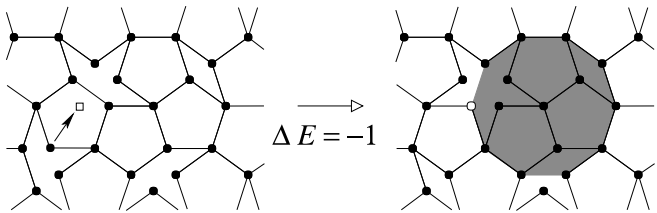}
\caption{Creation of a new cluster by a single flip, which lowers the
total energy.}
\label{fig_new_cluster}
\end{figure}

We have found (see Sec.~\ref{sec_entropy}) that the flip
moves shown in Fig.~\ref{fig_flips} have the required properties.  
The flip configurations consist of a hexagon with an interior
vertex which can jump to its ``mirror image''. This move corresponds to a
change in the orientation of the hexagon by $180^{\circ}$.
The vertices of the hexagon itself are not affected, but the adjacent tile
configurations are changed depending on the local environment. In
Fig.~\ref{fig_flips}(a), e.~g., the adjacent ship and rhombus are 
exchanged, or in Fig.~\ref{fig_flips}(b), a star and a rhombus are
transformed into two ships, etc. These flips preserve the
property that the spiky tiles are always surrounded by pentagons. 
There are some ``flip configurations'' where the local environment 
prohibits the flip. This is the case for configurations like in
Fig.~\ref{fig_flips}(d), where the flip would introduce a new kind of
tile (see also Ref.~\onlinecite{hon02}). In all our simulations, such
flips were forbidden. For the sake of completeness, it is explicitly shown in
Fig.~\ref{fig_new_cluster} how a new cluster can be created by a
single flip. 

\begin{figure}[t]
\includegraphics[width=\columnwidth]{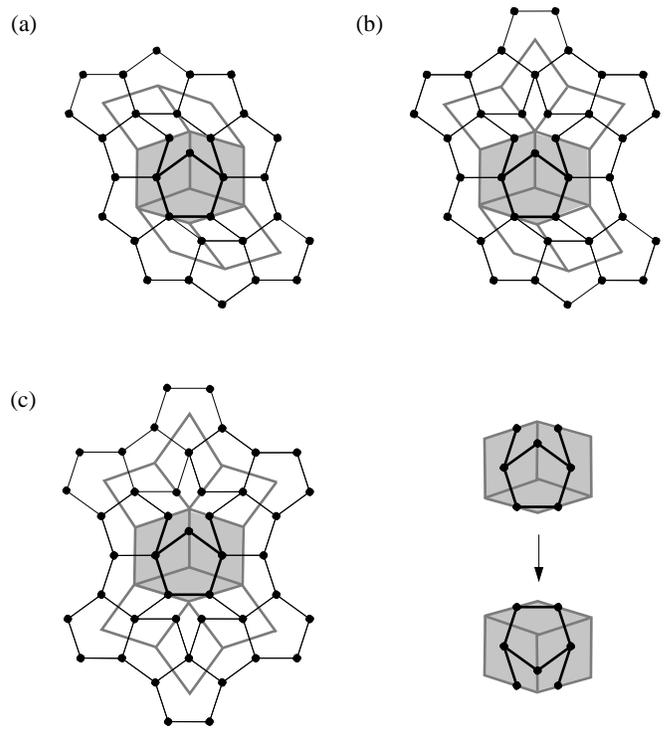}
\caption{Relationship between the flip configurations in the PPT
(black), as shown in Fig.~\ref{fig_flips}, and the corresponding local
tile configurations in the random Penrose rhombus tiling (gray), which
were used in Ref.~\onlinecite{tan90}. The flip move is
explicitly shown in the bottom right corner.}
\label{fig_hex-flips}
\end{figure}

The flips used for the PPT are in direct correspondence with
hexagon flips in the associated 4-level Penrose tiling. This is 
illustrated in Fig.~\ref{fig_hex-flips}. Note that the other type of
hexagon flip plays no role, because it leaves the HBS tiling, and
thus the PPT invariant. As it is known that hexagon flips are 
ergodic for the 4-level Penrose tiling, this correspondence
adds further confidence that the PPT flips are indeed ergodic, too.

With this MC scheme, the states of maximal cluster density 
can be determined by simulated annealing, using as energy 
the negative of the number of clusters, thus mimicking the total cohesion
energy of the clusters. Then the energetic ground state, reached at
low temperatures, is the ensemble of states with maximal cluster density.
The method we use is based on the Metropolis 
importance sampling algorithm.\cite{tan90,new99} The basic MC
move is as follows: (i)~Choose a vertex randomly. (ii) If it can 
be flipped, calculate the energy change $\Delta E$ (which is the
negative of the change in the number of clusters) and flip it with
probability
\begin{equation} 
\label{metropolis-algortihm}
p=
\left\{
\begin{array}{cl}
\e^{-\beta\Delta E} & \text{for }\Delta E>0 \,, \\
1 & \text{otherwise} \,.
\end{array}
\right.
\end{equation}
This algorithm fulfills the condition of detailed balance.

\begin{figure}[t]
\includegraphics[clip,width=\columnwidth]{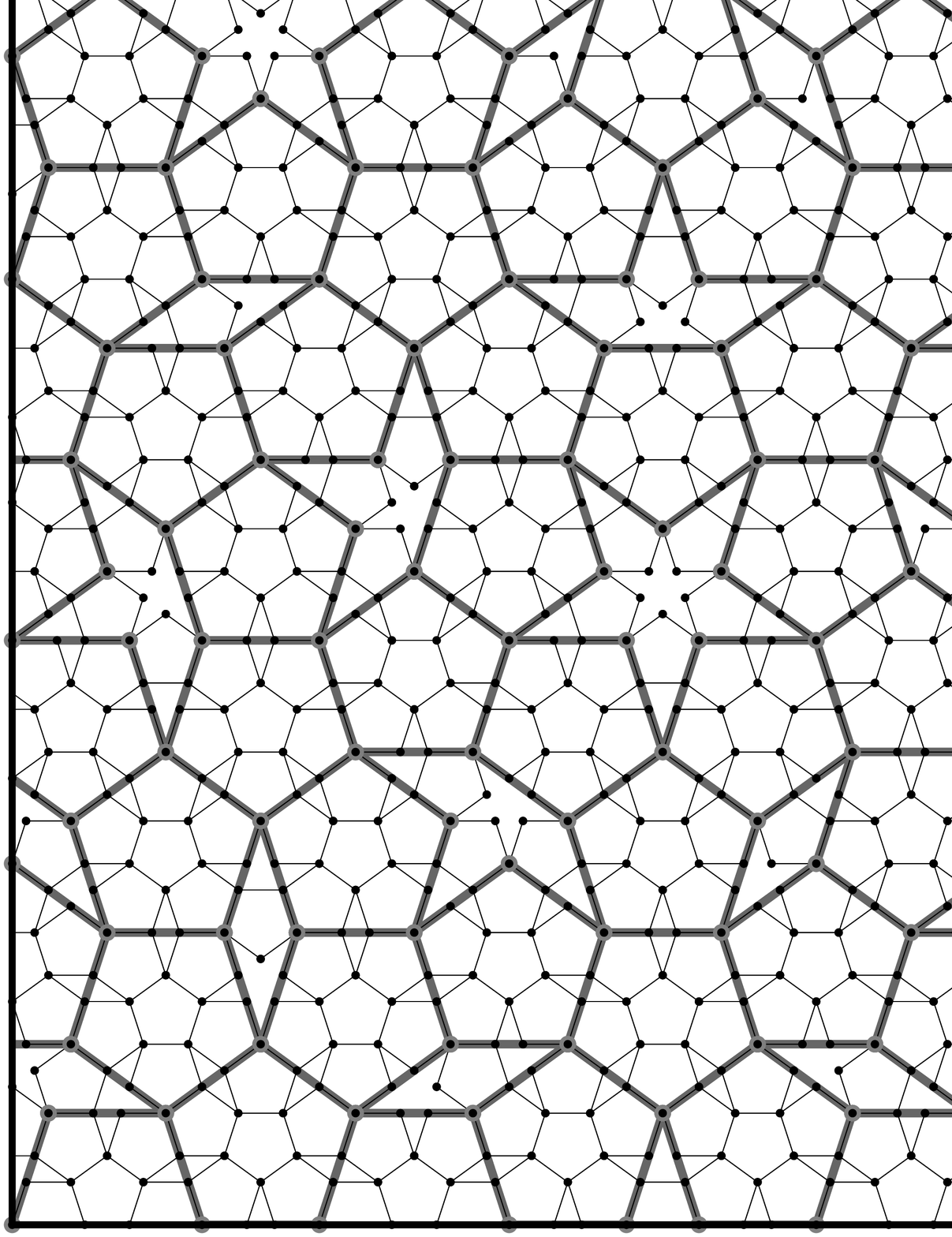}
\caption{Structure with maximal cluster density. The cluster centers form
the vertices of a supertile random PPT (gray lines).}
\label{fig_supertiling}
\end{figure}

It turns out that the states of maximal cluster density are {\it supertile 
random PPTs}, whose tiles have an edge length $\tau^{2}$ times that of the 
small tiles (where $\tau=(1+\sqrt{5})/2$ is the golden number). An
example of such a supertiling is shown in
Fig.~\ref{fig_supertiling}. It cannot be a perfect tiling, since it 
is still possible to move clusters in the ground state without
changing their number. This is due to the fact that the
relaxed rule and hence our vertex cluster does not enforce the
orientation of the A-overlaps.

In view of the results of the previous section, 
this is of course not too surprising. The cluster centers sit on the 
vertices of the supertiling, covering all vertices of the small 
tiles. Since the vertex cluster is smaller than the Gummelt decagon, 
it does not cover the whole area, but only the vertices; there remain 
small pentagons uncovered, which sit at the center of the supertile 
pentagons. This does not affect the overlap constraints, however.

Our results therefore imply that there is a one-to-one correspondence 
between {\it decagon coverings satisfying the relaxed rule} and
structures with {\it maximal density of the vertex cluster}
(Figs.~\ref{fig_covering} and \ref{fig_supertiling}, respectively). 
Although these two ordering principles are very similar, they are 
conceptually slightly different and have to be distinguished.

Supertile random tiling ensembles as a result of cluster maximization
were found already in Ref.~\onlinecite{jeo94}. Maximizing star decagon 
clusters in a rhombus tiling leads to an ensemble of HBS-type supertile 
tilings.\cite{jeo94,hen97} That ensemble contains also other structures, 
however, which is not the case for our cluster. In particular, it should 
be noted that if the tile stoichiometry of the pentagons and spiky tiles 
admits an HBS supertile tiling, then the state of maximal cluster density 
is always an HBS supertile tiling. A phase separation as discussed in 
Ref.~\onlinecite{hen97} is not possible. The only tiles that could be 
separated are the thin rhombi, but if there are enough pentagons, 
it is always advantageous to surround the rhombi with pentagons. 
The ensemble obtained by maximizing our vertex cluster is therefore 
a strict subensemble of the one obtained by maximizing the star decagon 
in random rhombus tilings.\cite{hen97}

\section{Entropy density}
\label{sec_entropy}

\begin{figure}[t]
\includegraphics[width=\columnwidth]{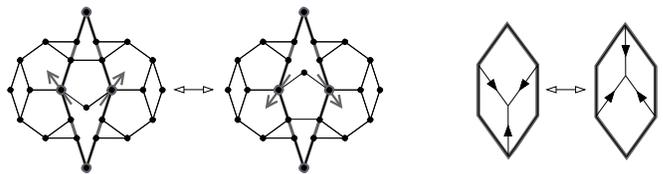}
\caption{In the supertile random PPT, each rhombus is counted twice because
of the two possible cluster orientations on the obtuse corners (left).
This is in one-to-one correspondence with the hexagons in the
4-level random tiling (right).}
\label{fig_double-count}
\end{figure}

With our cluster model, it is also possible 
to measure the entropy density of the ensemble of structures with
maximal cluster density and thus the entropy density of the {\it relaxed
cluster covering ensemble}. In the previous section, we have
introduced an energy model which assigns a cohesion energy to each
cluster in the structure. In this model, the ground state, i.~e., the
state of maximal cluster density, consists of supertile random PPTs
with an extra weight of two per rhombus, because for each rhombus
there are two choices of a cluster configuration with the same number
of clusters, as shown in Fig.~\ref{fig_double-count}
(see also Fig.~\ref{fig_orientations_2}(e)). 
At infinite temperature, on the other hand,
we have the full random PPT (at the level of the small tiles), where
each rhombus is counted only once.

With the entropic sampling algorithm,\cite{lee93,new99} we can determine
the entropy of the system as a function of energy. In this method, the
Boltzmann probability $\e^{-\beta E}$ of the Metropolis
algorithm is replaced by the factor $\e^{-S(E)}$, where
$S(E)$ is the microcanonical entropy function. Since this factor is
just the inverse of the number $g(E)$ of states with energy $E$, according to
\begin{equation}
\label{entropy}
S(E)=\ln g(E) 
\end{equation}
(in units of $\kB$), we obtain a uniform energy distribution
\begin{equation}
P(E)\propto g(E)\e^{-S(E)}\equiv 1 \,,
\end{equation} 
which corresponds to a random walk through the energy space of the system.

However, the exact entropy function $S(E)$ is not known a priori, hence it
has to be determined iteratively, starting with a rough estimate of
$S(E)$. This estimate can be obtained by a short run using as
``entropy function'' $S(E)\equiv 0$, which yields an estimate for the
degeneracy $g(E)$ of the different energies and thus, via
Eq.~(\ref{entropy}), an estimate for 
the real entropy function. Another possibility to get a good estimate is to
take advantage of the extensive nature of entropy by scaling the
entropy function of a smaller system to a larger one.

Subsequently, the entropy function is optimized by an iterative
procedure. Analogously to the Metropolis algorithm, we choose for
the flip probabilities  
\begin{equation} 
\label{entropic-sampling}
p=
\left\{
\begin{array}{cl}
\e^{-\Delta S} & \text{for }\Delta S>0 \,, \\
1 & \text{otherwise} \,,
\end{array}
\right.
\end{equation}
where $\Delta S=S(E+\Delta E)-S(E)$ is the change in entropy due to
the considered MC move (with energy change $\Delta E$). This choice of $p$
likewise fulfills the condition of detailed balance.
The iteration scheme is then as follows: (i) Run a MC simulation based on
Eq.~(\ref{entropic-sampling}) in order to obtain an energy histogram
$H(E)$. (ii) Use the measured histogram to correct the entropy
function according to the update rule
\begin{equation}
S(E)\leftarrow
\left\{
\begin{array}{ll}
S(E)+\ln H(E) & \text{for }H(E)\neq 0 \,,\\
S(E) & \text{for }H(E)=0 \,.
\end{array}
\right.
\end{equation}
(iii) Continue at (i) until the energy histogram is sufficiently uniform. 

\begin{figure}[t]
\includegraphics[width=\columnwidth]{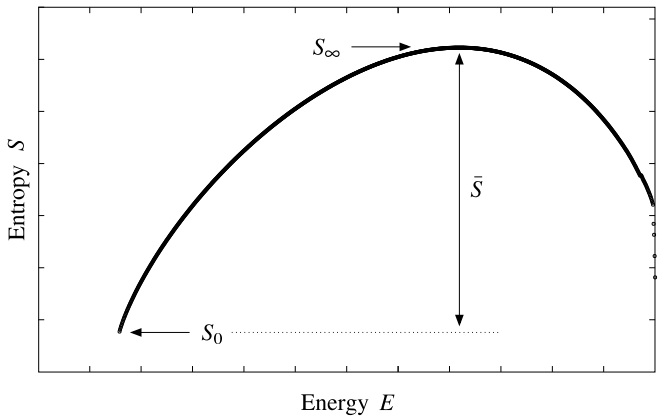}
\caption{Representative example of the microcanonical entropy as a
function of the energy of the system.}
\label{fig_entropy-function}
\end{figure}

An example of such an entropy function is shown in
Fig.~\ref{fig_entropy-function}. As this method does not yield
absolute entropy values, we can only measure entropy differences, in
particular the difference $\bar{S}$ between the ground state and the 
infinite-temperature state (i.~e., the maximum of the entropy function). 
The entropies at zero ($S_0$) and infinite temperature ($S_\infty$) 
are both entropies of random PPTs, once with a twofold degeneracy 
for each rhombus (Fig.~\ref{fig_double-count}), and once without.
To compare these two entropy values, the latter one has to be corrected
by adding an extra double-counting of the rhombi, which yields for 
each rhombus a factor of 2 in the degeneracy $g_\infty$ or an additive
contribution of $\ln 2$ in the entropy $S_\infty$, respectively.   
Moreover, the two random tilings are at different length scales, since
the supertile edges are $\tau^2$ times larger than the small tile
edges, which has to be taken into account due to the 
extensive nature of entropy. Therefore, we have the following
relation between the two entropy densities $\sigma_0$ and $\sigma_\infty$:
\begin{equation}
\tau^4\sigma_0=\sigma_\infty+\rhorh\ln 2 \,,
\end{equation}
where $\rhorh$ is the measured rhombus density in the infinite-temperature
state. If we write $\sigma_\infty=\sigma_0+\bar{\sigma}$, 
we end up with an equation for the ground state entropy density $\sigma_0$, 
in which all other quantities can be measured:
\begin{equation}
\sigma_0=\frac{1}{\tau^4-1}(\bar{\sigma}+\rhorh\ln 2) \,. 
\end{equation}

\begin{figure}[t]
\includegraphics[width=\columnwidth]{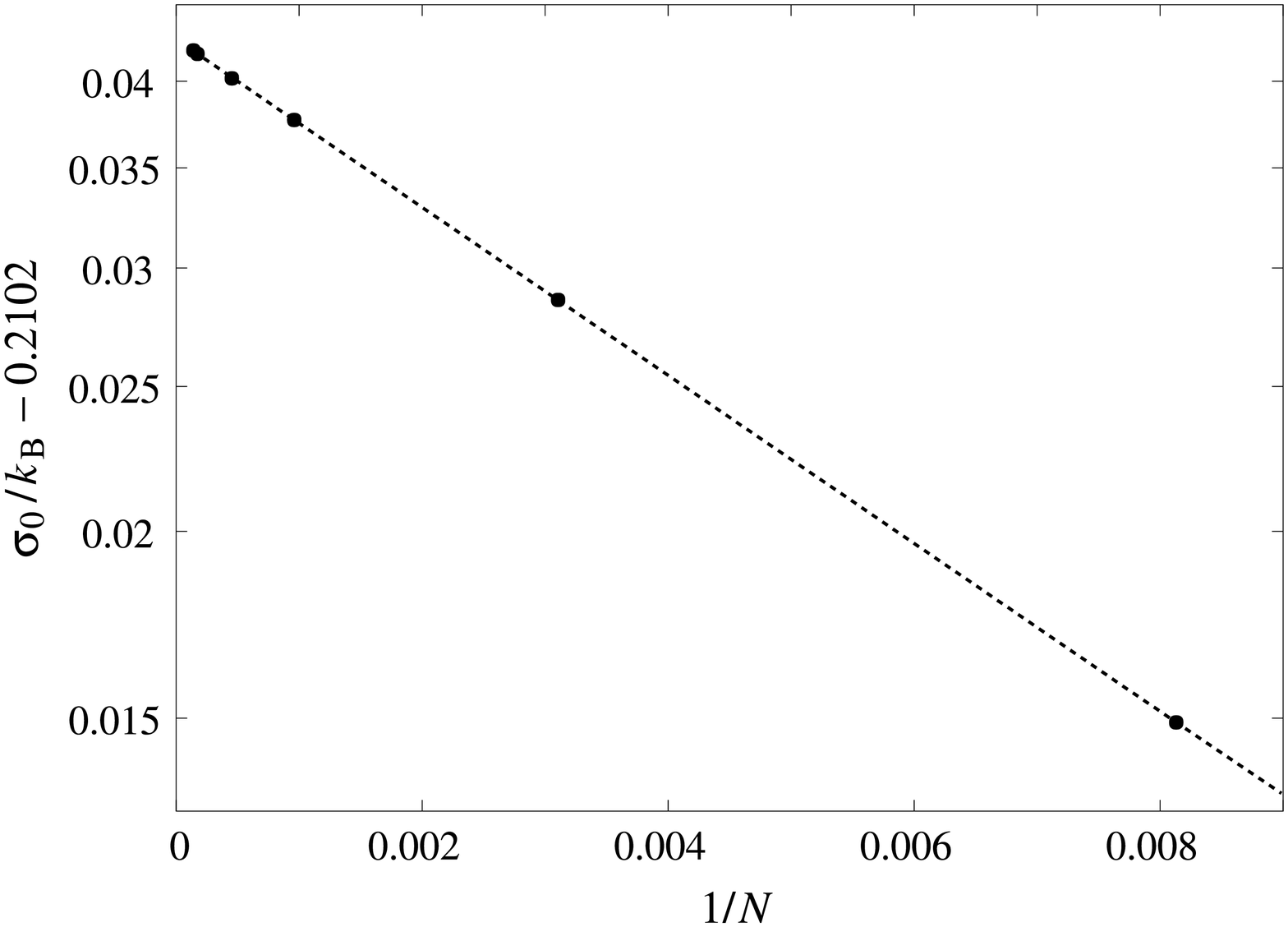}
\caption{Finite-size scaling of the entropy density of the relaxed
cluster coverings, where $N$ is the number of vertices.
The applied fit function is of the (empirical) form
$\sigma(N)=a+b\e^{-c/N}$. In this graph, the line
$\log[\sigma(N)-a]=\log b-c/N$ is shown. The error bars are smaller
than the plot symbols.}
\label{fig_entropy-scaling}
\end{figure}

The ground state entropy density has been determined in this way for 
several periodic approximants. By finite-size scaling, the values can 
then be extrapolated to infinite system size. For this purpose, the entropy
density, as a function of system size $N$, is fitted to a function
of the (empirical) form $a+b\e^{-c/N}$, which proves to work very well
(Fig.~\ref{fig_entropy-scaling}). At a scale where the supertile 
edges (which separate A-overlap neighbors in the cluster model) have 
unit length, we obtain a value of 
\begin{equation}
\sigma_0/\kB=0.253\pm 0.001
\end{equation}
for the {\it entropy density of the relaxed coverings}. 
This can be compared with the value which Tang and Jari\'c have
obtained by Metropolis-type MC simulations for the entropy density of
the 4-level random tiling.\cite{tan90} In Sec.~\ref{sec_covering}, we
have seen that 4-level random tilings are in one-to-one correspondence
with relaxed cluster coverings (see also
Fig.~\ref{fig_double-count}). If the different length scales of the 
two tilings are taken into account by a simple geometric conversion,
the value of Tang and Jari\'c turns into
\begin{equation}
\sigma_0/\kB=0.255\pm 0.001 \,,
\end{equation}
which is compatible with our result.

\section{Coupling between clusters}
\label{sec_coupling}

\begin{figure}[t]
\includegraphics[width=\columnwidth]{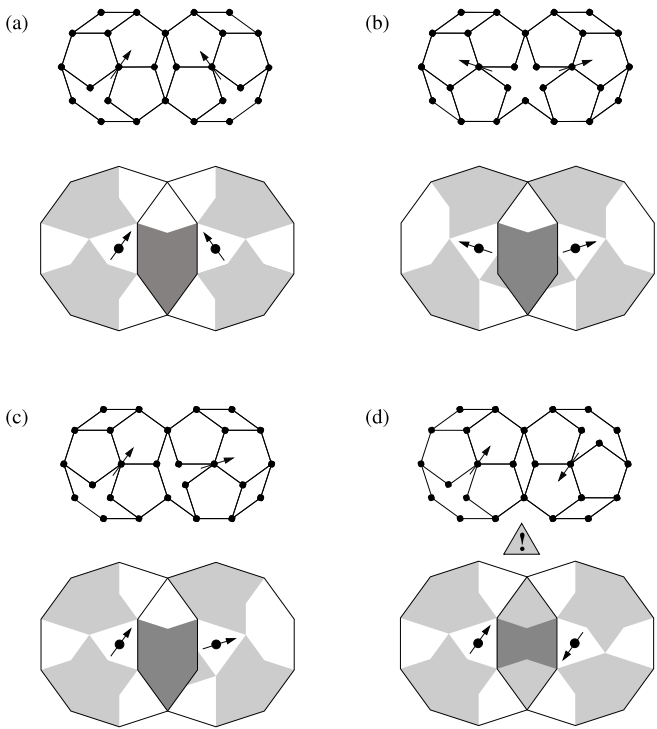}
\caption{Possible A-overlaps in supertile random PPTs or relaxed
coverings. (a--c) obey the perfect rule. Only (d) is a disoriented
overlap.}
\label{fig_a-overlaps}
\end{figure}

The only difference between the {\it perfect} and the {\it relaxed}
overlap rule is that the latter does {\it not} require {\it oriented}
A-overlaps (whereas the orientation of the B-overlaps has to be
obeyed). Since not all relaxed coverings are perfect, 
there must be A-overlaps which do {\it not} obey the
orientation condition of the perfect rule. A closer analysis shows
\cite{gum00,gum02} that there is actually only one kind of disoriented
A-overlap. All A-overlaps which can occur in relaxed coverings or
supertile random PPTs, respectively, are shown in Fig.~\ref{fig_a-overlaps}. 
For the disoriented A-overlap not permitted by the perfect rule
(Fig.~\ref{fig_a-overlaps}(d)), the two clusters have antiparallel
orientations.

To order the (supertile) random tiling structures to perfect tilings, 
we introduce a coupling between neighboring
clusters in such a way that overlaps which are not permitted by the
perfect rule are energetically penalized. We expect such a coupling
to be weak, because these kinds of defects can be detected only
at larger scales. However, at low temperatures this coupling might still
be able to order the supertile random tiling ground state of the
relaxed cluster covering to a perfectly quasiperiodic structure. 

\begin{figure}[t]
\includegraphics[width=\columnwidth]{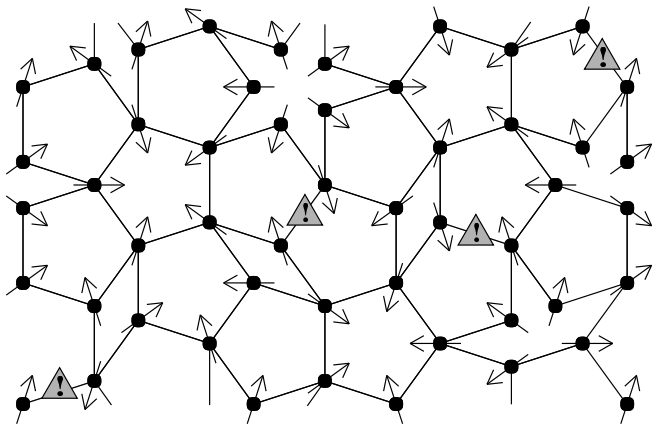}
\caption{Supertiling with cluster orientations indicated by arrows.
The forbidden A-overlaps, corresponding to tile edges with
antiparallel arrows at their ends, are marked.}
\label{fig_arrow-model}
\end{figure}

This suggests a scenario with two energy or temperature scales.
The presence of each vertex cluster 
lowers the cohesion energy by a large amount, so that structures
with maximal cluster density are strongly favored, even at relatively
high temperatures. The equilibrium structures at these temperatures 
are therefore relaxed cluster coverings. Additionally, there is 
a small coupling between neighboring clusters, which can order the
supertile random tiling to a perfect tiling at low temperatures.

We have verified the feasibility of this approach by MC simulations.
For this purpose, we consider only the subensemble of states with
maximal cluster density. In other words: We run our simulations at
the level of the supertiling. If we represent a cluster in the
supertile PPT by its center and its orientation (given by an arrow),
we can describe the covering in a much more compact way, as shown in
Fig.~\ref{fig_arrow-model}. Such a setup keeps the number of clusters 
constant, so that we cannot leave the states of maximal cluster density, 
which simplifies the simulation considerably. 

\begin{figure}[t]
\includegraphics[width=\columnwidth]{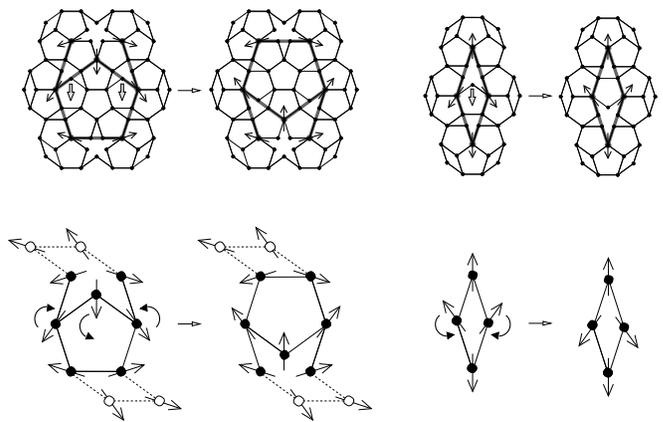}
\caption{Basic flips in the underlying tiling (top) and effective
moves on the level of the supertiling (bottom). The
hexagon flip (left) is the same as the one used before. 
Additionally, we also have the rhombus flip (right) which
does not change the vertex structure of the supertiling. For both flip types,
the cluster orientations have to be adjusted consistently with the
underlying structure.}
\label{fig_arrow-flips}
\end{figure}

\begin{figure}[t]
\includegraphics[width=\columnwidth]{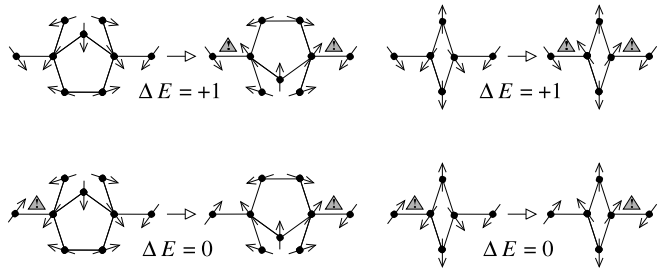}
\caption{Creation (top) and shift of defects (bottom) by single flips
of the two types: hexagon flip (left) and rhombus flip (right).}
\label{fig_defect-moves}
\end{figure}

As flip moves we can still
use those of Fig.~\ref{fig_flips}, except that we now have to adjust the 
cluster orientations of several vertices in the flip configuration. We
have to do this in a way consistent with the vertex configuration in the 
underlying tiling, which actually determines the cluster orientations 
(Fig.~\ref{fig_arrow-flips}). The first type of flip consists of two 
simultaneous flips in the underlying tiling. We have to point out that 
for a hexagon-type flip configuration like the one in 
Fig.~\ref{fig_arrow-flips} (bottom left) the
cluster orientations on the obtuse corners of the adjacent rhombus
have to be as shown. Otherwise, the flip is not allowed, since then
the cluster orientations would be inconsistent with the ship-shaped vertex
configuration after the flip.

Additionally, we have to introduce a new type of flip which only changes the 
cluster orientations on the obtuse corners of a rhombus, but keeps the
tiling itself fixed. Such a move corresponds to a single flip in the
underlying tiling. In comparison with the flip moves in a
Penrose rhombus tiling,\cite{hen91,tan90} the hexagon flip
corresponds to D-type and the rhombus flip to Q-type
configurations.\cite{bru81}  

This MC dynamics can change the number of defects, as demonstrated in
Fig.~\ref{fig_defect-moves}. Again, we have used a Metropolis-type MC
scheme,\cite{tan90,new99} now with the number of defects as energy. 
We have seen in our simulations that the coupling of the
clusters, which penalizes the defects, is indeed capable of ordering
the random tilings to perfectly quasiperiodic structures. 
In other words: The ground state, reached by simulated annealing, is a
perfect PPT, whereas the high-temperature state is a supertile random
tiling or relaxed cluster covering. In this
respect, ``high'' temperature means high compared to the cluster coupling,
but still low compared to the energy required to break up clusters.

We have to mention that the ground state is not always a perfect PPT.
Perfectness is determined by lifting the vertices to hyper-space; 
if the projections of the vertices onto the perpendicular space are 
all inside the acceptance region, the tiling is perfect, otherwise 
it is not. Since we use periodic approximants of PPTs in our 
simulations, the minimal number of defects is always larger than 
zero.\cite{ent88} These defects can be shifted without changing 
the energy. In this process, a vertex can leave the acceptance region 
in the perpendicular space. Thus, the tilings with minimal number of
defects are not necessarily perfect. 

With the model of
coupled clusters, it is also possible to measure the entropy density of the
relaxed covering ensemble. In this case, the ground state is ordered
and has zero entropy (at least in the thermodynamic limit), and the
high-temperature state is the one whose entropy we are interested in.
We therefore only need to measure the difference between the entropies of
the high-temperature state and the ground state, and extrapolate
these values to infinite system size. It turns out, however, that the
finite-size scaling for this model does not work as well as for the model
with a supertile random tiling in the ground state as considered in 
Sec.~\ref{sec_entropy}. The results are therefore less precise, but 
nevertheless consistent.

\section{Discussion and conclusion}
\label{sec_conclusion}

In this paper, we have discussed different ordering principles for 
quasicrystals
based on the cluster picture, namely the cluster covering principle and
the principle of cluster density maximization. A relaxed version  of
the covering rules for Gummelt's aperiodic decagon has been considered,
which produces as ground state a supertile random PPT. It has been shown
that this relaxed overlap rule has a very natural realization in terms
of a vertex cluster in the PPT. 

The feasibility of our model has been tested by MC
simulations. In particular, we have verified that the
relaxed cluster coverings coincide with the states of maximal
cluster density. The entropy density measured in the random 
covering ensemble is found to agree with the entropy density 
obtained by Tang and Jari\'c for the equivalent random 4-level tiling.
Moreover, we have shown that a coupling between neighboring clusters
can order the random-tiling-type ground states to a perfectly
quasiperiodic structure.

Models of this kind can be very suitable for the explanation of 
experimentally observed decagonal quasicrystals, as the latter are often not 
perfectly quasiperiodic, but more of a random tiling nature.\cite{gum02} 
Random PPTs, and the closely related 
HBS tilings, are often observed in high-resolution electron microscopy 
images (Fig.~\ref{fig_em-picture}, compare Figs.~\ref{fig_covering}
and \ref{fig_supertiling}).\cite{rit96,sol01} 

\begin{figure}[t]
\centerline{\includegraphics[width=7.5cm]{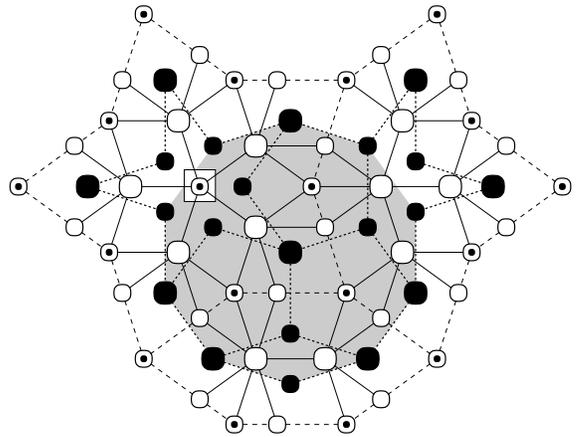}}
\caption{Atomic cluster found by Roth and Henley \cite{rot97} in a molecular 
dynamics simulation. White atoms are at $z=0$, black atoms at $z=\frac12$, 
and dotted atoms at $z=\frac14,\frac34$ (in units of the period in 
$z$-direction). The area of our vertex cluster is shaded in gray.
At the center of a star (marked with a box), we would
have expected a different atomic configuration (see text).}
\label{fig_roth-henley-cluster}
\end{figure}

The most striking resemblence, however, is with a three-dimensional
atomic cluster found by Roth and Henley \cite{rot97} in a molecular 
dynamics simulation of binary decagonal Frank-Kasper-type quasicrystals 
(Fig.~\ref{fig_roth-henley-cluster}), whose lateral overlap constraints
are almost the same as those of our two-dimensional vertex cluster. 
The only discrepancy is at the center of a star, where for full
equivalence we would have expected a single atom at $z=0$ in the 
Roth-Henley cluster, not two atoms at  $z=\frac14,\frac34$.

The vertex set of our cluster can be realized by several tile
configurations in the PPT (Fig.~\ref{fig_cluster}). This might
possibly be related to the results of an Al-Co-Ni structure analysis by
Cervellino, Haibach, and Steurer.\cite{cer02} They observed that
there is a perfectly quasiperiodic long-range order of the centers of
atomic clusters, but the interior structure of the clusters themselves
is disordered. This means that the local atomic interactions cannot
enforce local order, but there exists a long-range order.
The proposed mechanism for long-range ordering is the electronic
long-range term of free electrons. These long-wavelength electrons
``see'' only a simpler, ``smeared out'' version of the complex cluster
structure.

As we have seen in Sec.~\ref{sec_coupling}, the shift of a flipping
cluster center is large, whereas the corresponding two vertex moves in
the underlying tiling are small (Fig.~\ref{fig_arrow-flips}). 
In the atomic simulations of Al-Co-Ni done by Honal and
Wellberry,\cite{hon02} the flip of a cluster corresponds to jumps of only
four atoms. The magnitude of these jumps is small (about 1~\AA)
compared to the scale of the clusters (about 20~\AA). 

All our simulations have been in two dimensions. It is well known that, for
finite-range interactions, quasicrystal structures in two dimensions cannot be
stable at any positive temperature.\cite{kal89} At non-zero temperatures,
they are always in a random tiling state, the ``phase transition'' from the
ordered phase to the random tiling state being at zero temperature. 
However, in three dimensions the critical temperature is expected to
be positive. For this purpose, one can consider as three-dimensional
system a stacking of our two-dimensional models with a
suitable coupling between neighboring layers. Such layered {\it tiling} models 
have already been studied by Jeong and Steinhardt,\cite{jeo93} but are
possible also in the {\it cluster covering} setting. As expected
for three-dimensional systems, the phase transition from
ordered to random-tiling-type structures is at finite
temperature. The results of these studies will be presented in a
separate paper.\cite{part2}

\begin{acknowledgments}
We would like to thank Petra Gummelt for useful discussions on the
relationships between the different relaxed coverings and random
tiling ensembles.
\end{acknowledgments}

\end{document}